\documentclass[fleqn,usenatbib]{mnras}
\usepackage{mathrsfs}
\usepackage{amssymb,amsmath}
\usepackage{hyperref}
\usepackage{newtxtext,newtxmath}
\usepackage[T1]{fontenc}
\usepackage{ae,aecompl}
\usepackage{graphicx}	
\usepackage{amsmath}	
\usepackage{amssymb}	

\DeclareMathAlphabet{\mathbi}{OT1}{ptm}{bx}{it}
\SetMathAlphabet\mathbi{bold}{OT1}{ptm}{bx}{it}

\title[A New Approach for Measuring Power Spectra in AGNs]{A New Approach for Measuring Power Spectra and Reconstructing Time Series in Active Galactic Nuclei}

\author[Y.-R. Li \& J.-M. Wang]{
  Yan-Rong Li$^{1}$\thanks{E-mail: \href{mailto: liyanrong@mail.ihep.ac.cn}{liyanrong@mail.ihep.ac.cn}} and Jian-Min Wang$^{1,2,3}$\\
$^1$Key Laboratory for Particle Astrophysics, Institute of High 
Energy Physics, Chinese Academy of Sciences,\\ 19B Yuquan Road, 
Beijing 100049, China\\
$^2$School of Astronomy and Space Science, School of Physical Sciences, University of Chinese
Academy of Sciences, \\
19A Yuquan Road, Beijing 100049, China\\
$^3$National Astronomical Observatories of China, Chinese Academy of Sciences, 20A Datun Road,
Beijing 100020, China
}

\date{Accepted XXX. Received YYY; in original form ZZZ}

\pubyear{2017}

\begin{document}

\label{firstpage}
\pagerange{\pageref{firstpage}--\pageref{lastpage}}
\maketitle

\begin{abstract}
We provide a new approach to measure power spectra and reconstruct time series in active galactic nuclei (AGNs) based on 
the fact that the Fourier transform of AGN stochastic variations is a series of complex Gaussian random variables.
The approach parameterizes a stochastic series in frequency domain and transforms it back to time domain to fit the observed 
data. The parameters and their uncertainties are derived in a Bayesian framework, which also allows us to compare 
the relative merits of different power spectral density models. The well-developed fast Fourier transform algorithm 
together with parallel computation enable an acceptable time complexity for the approach.
\end{abstract}

\section{Introduction}
Active galactic nuclei (AGNs) are long known to vary in fluxes at all wavebands with a broad range
of time scales (e.g. see \citealt{Ulrich1997} for a review). The physical process responsible for AGN 
variability remains unclear. It is most likely that accretion disks surrounding the central black holes
are the dominant source for AGN activity and variability (\citealt{Shakura1973}). With current
and forthcoming large time-domain surveys such as the Catalina Real-Time Transient Survey (\citealt{Drak2009}), 
the All-Sky Automated Survey for Supernovae (\citealt{Kozlowski2017a}), and
the Large Synoptic Survey Telescope (\citealt{Ivezic2008}), we has entered in a new era for AGN variability studies.
The unprecedented amount of variability data will provide deep insight into the exact processes underlying 
AGN variability.

A standard method for characterizing AGN variability is using the power spectral density (PSD). However, deriving
the PSD and the associated uncertainties from an observed time series is always challenging because the effects such 
as red-noise leakage and aliasing imposed by the sampling pattern can easily distort the power spectrum from the true spectrum 
(\citealt{Uttley2002}). Great efforts have been made to account for spectral distortion in different levels (e.g., 
\citealt{Done1992, Uttley2002, Kelly2009, Kelly2014, Zhu2016, Kozlowski2017a}). In this letter, we propose an alternative 
method to measure PSDs in a fully Bayesian framework, which allows us to compare the relative merits of different
PSD models and determine the most probable one. Meanwhile, the method simultaneously 
provides stochastic reconstructions to an observed time series, which are useful for otherwise studies such as 
reverberation mapping analysis (\citealt{Li2013, Li2014, Pancoast2014}).

\section{Methodology}

\begin{figure*}
\centering
\includegraphics[width=0.8\textwidth, trim=0pt 10pt 0pt 10pt]{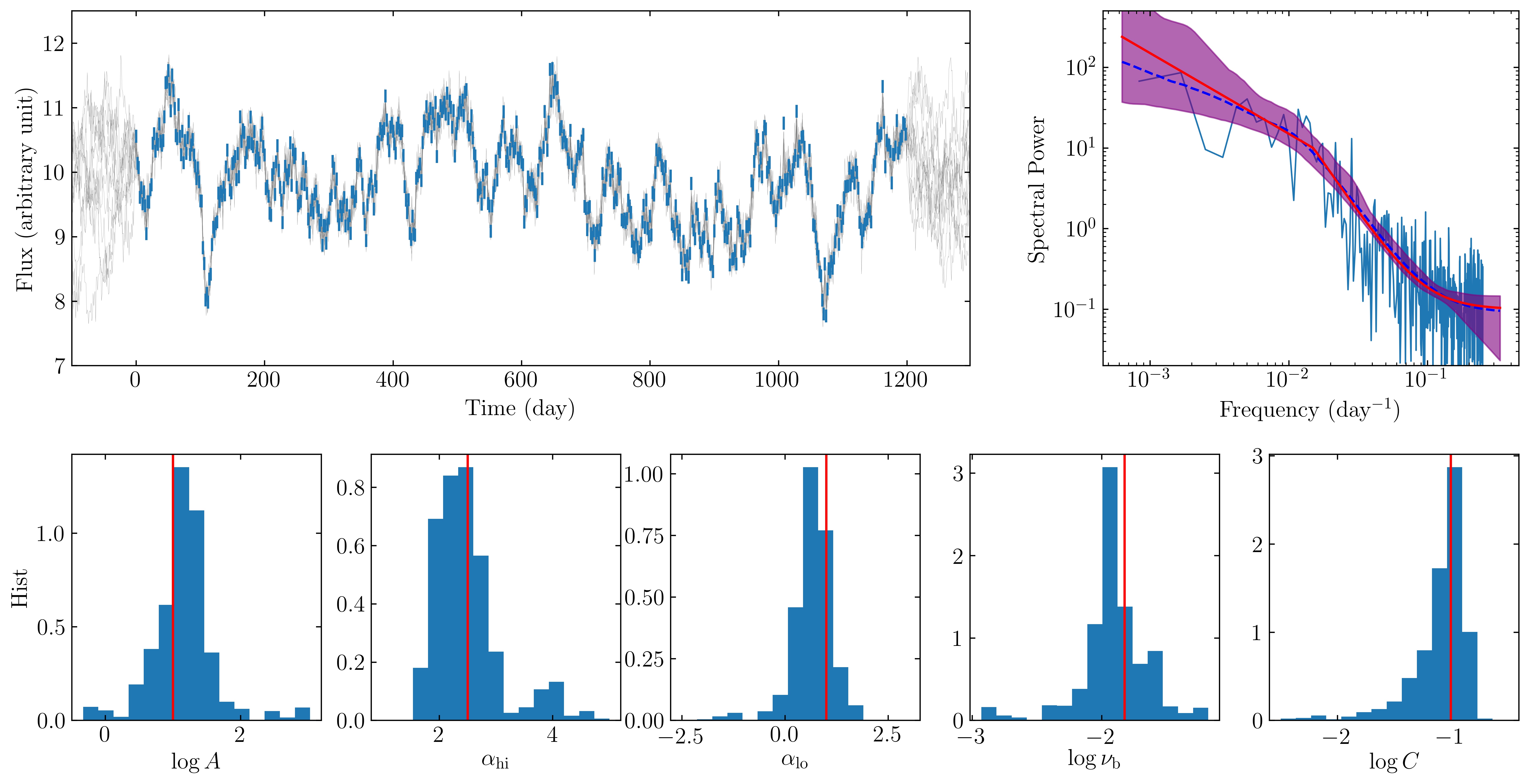}
\caption{Validity test of our approach. (Top left) Blue points with errorbars represent a mock light curve with a bending power-law PSD 
(equation \ref{eqn_bending_pow}). 
Thin grey lines represent stochastic reconstructions. (Top right) The periodogram 
of the mock light curve. Red solid line represents the input PSD and blue dashed line represents the best inferred PSD.
Shaded area represents the 2$\sigma$ error bands (95\% confidence level). (Bottom) Posterior distributions of the PSD parameters. Red solid line 
represent the input values.}
\label{fig_sim}
\end{figure*}

\subsection{Mathematical Preliminaries}
The (discrete) Fourier transform of a time series $\mathbi{x}$ is defined as (e.g., \citealt{Timmer1995})
\begin{eqnarray}
\mathbi{f}(\nu) =  \mathscr{F}(\mathbi{x}) = \frac{1}{\sqrt{N}}\left[\sum_t x(t) \cos(2\pi\nu t) - i\sum_t x(t) \sin(2\pi\nu t)\right],
\label{equ_ft}
\end{eqnarray}
where $i$ is the imaginary unit, $\nu$ is frequency, and $N$ is the number of points.
As time series that we cope with is generally real, $\mathbi{f}$ has a property of $f(-\nu)=f^*(\nu)$, i.e., the component at negative frequency
is equal to the complex conjugate of the component at the corresponding positive frequency.
The inverse (discrete) Fourier transform is the inverse operation of equation (\ref{equ_ft}), given by
\begin{eqnarray}
\mathbi{x}(t) = \mathscr{F}^{-1}(\mathbi{f}) =\frac{1}{\sqrt{N}}
\left[\sum_{\nu} f(\nu) \cos(2\pi\nu t) + i\sum_\nu f(\nu) \sin(2\pi\nu t)\right].
\end{eqnarray}
From the theory of spectral estimation (\citealt{Priestley1981}), it is known that $f(\nu)$ 
can be expressed into complex Gaussian random variables
\begin{eqnarray}
\mathbi{f}(\nu|\boldsymbol{\theta}) = \sqrt{\frac{S(\nu|\boldsymbol{\theta})}{2}}\left[\mathscr{N}(0, 1) + i\mathscr{N}(0, 1)\right],
\label{eqn_freq}
\end{eqnarray}
where $\mathscr{N}(0, 1)$ is the standard normal distribution and $S(\nu|\boldsymbol{\theta})$ is the power spectral density which is described by a 
parameter set $\boldsymbol{\theta}$.

For a given PSD model $S$ and observation data $D$, the parameter estimation is obtained by exploring the posterior 
probability distribution $P(\boldsymbol{\theta}|D, S)$, 
which is related to the likelihood probability distribution $P(D|\boldsymbol{\theta}, S)$ through the Bayes' theorem,
\begin{eqnarray}
P(\boldsymbol{\theta}|D, S) = \frac{P(D|\boldsymbol{\theta}, S)P(\boldsymbol{\theta})}{P(D|S)}.
\label{eqn_bayes}
\end{eqnarray}
Here $P(\boldsymbol{\theta})$ is the probability distribution for the parameter $\boldsymbol{\theta}$ and $P(D|S)$ is the marginal
likelihood probability distribution which is also called evidence in light of its crucial role in model selection
(\citealt{Sivia2006}). $P(D|S)$ is obtained by marginalizing the likelihood probability over the prior probability 
for the parameters
\begin{eqnarray}
P(D|S) = \int P(D|\boldsymbol{\theta}, S) P(\boldsymbol{\theta})\rm d \boldsymbol{\theta}.
\label{eqn_evidence}
\end{eqnarray}

Suppose that we have two PSD models $S_1$ and $S_2$, the Bayes factor, defined by the ratio of the posterior probabilities for $S_1$ and $S_2$,
quantifies the relative merit of the two models (\citealt{Sivia2006}),
\begin{eqnarray}
K = \frac{P(S_1|D)}{P(S_2|D)} = \frac{P(D|S_1)}{P(D|S_2)} \frac{P(S_1)}{P(S_2)},
\end{eqnarray}
where $P(S_1)$ and $P(S_2)$ are the prior probabilities for the two models, respectively. 
Generally, we assign equal priors for $S_1$ and $S_2$. As a result, the Bayes factor is simply
the ratio of the evidence\footnote{However, there are limitations of using Bayes factor due to its possible (in some cases)
dependence on the choice of priors (see \citealt{Gelman2004}, Ch. 6).}
\begin{eqnarray}
K = \frac{P(D|S_1)}{P(D|S_2)}.
\end{eqnarray}

\subsection{Bayesian Inference}
A realization of an observed time series $\mathbi{y}$ can be deemed into the sum of an underlying signal $\mathbi{x}$ and
a measurement noise $\mathbi{n}$, namely, 
\begin{eqnarray}
\mathbi{y}=\mathbi{x}+\mathbi{n}.
\end{eqnarray}
In frequency domain, the signal $\mathbi{x}$ corresponds to its Fourier transform $\mathbi{f}$ with a PSD $S$ described by
the parameters $\boldsymbol{\theta}$.
Assuming that the measurement noise $\mathbi{n}$ is Gaussian and uncorrelated, 
the likelihood probability distribution for $\mathbi{y}$ given $\mathbi{f}$ and the parameters $\boldsymbol{\theta}$ is 
\begin{eqnarray}
P(\mathbi{y}|\mathbi{f}, \boldsymbol{\theta}, S) & = & P(\mathbi{n} = \mathbi{y} - \mathbi{x}|\boldsymbol{\theta}, S)\nonumber\\
& = & P[\mathbi{n} = \mathbi{y} - \mathscr{F}^{-1}(\mathbi{f}|\boldsymbol{\theta}, S)]\nonumber\\
& = & \prod_j \frac{1}{\sqrt{2\pi} \sigma_j} \exp\left\{-\frac{[y_j - \mathscr{F}^{-1}(\mathbi{f}|\boldsymbol{\theta}, S)_j]^2}{2\sigma_j^2}\right\},
\end{eqnarray}
where $\sigma_j$ is the measurement noise of the $j$th point in the time series.
Using equation (\ref{eqn_bayes}), the posterior probability distribution for $\mathbi{f}$ and $\boldsymbol{\theta}$ 
is 
\begin{eqnarray}
P(\mathbi{f}, \boldsymbol{\theta}|\mathbi{y}, S) = 
\frac{P(\mathbi{y}|\mathbi{f}, \boldsymbol{\theta}, S)P(\mathbi{f}, \boldsymbol{\theta}|S)}{P(\mathbi{y}|S)} \nonumber\\
\propto P(\mathbi{y}|\mathbi{f}, \boldsymbol{\theta}, S)P(\mathbi{f}, \boldsymbol{\theta}|S),
\label{eqn_post}
\end{eqnarray}
where $P(\mathbi{f}, \boldsymbol{\theta}|S)$ is the prior probability 
distribution for $\mathbi{f}$ and $\boldsymbol{\theta}$ and the  marginal likelihood probability $P(\mathbi{y}|S)$ is
constant for given $S$.

\begin{figure}
\centering
\includegraphics[width=0.45\textwidth, trim=0pt 0pt 0pt 0pt]{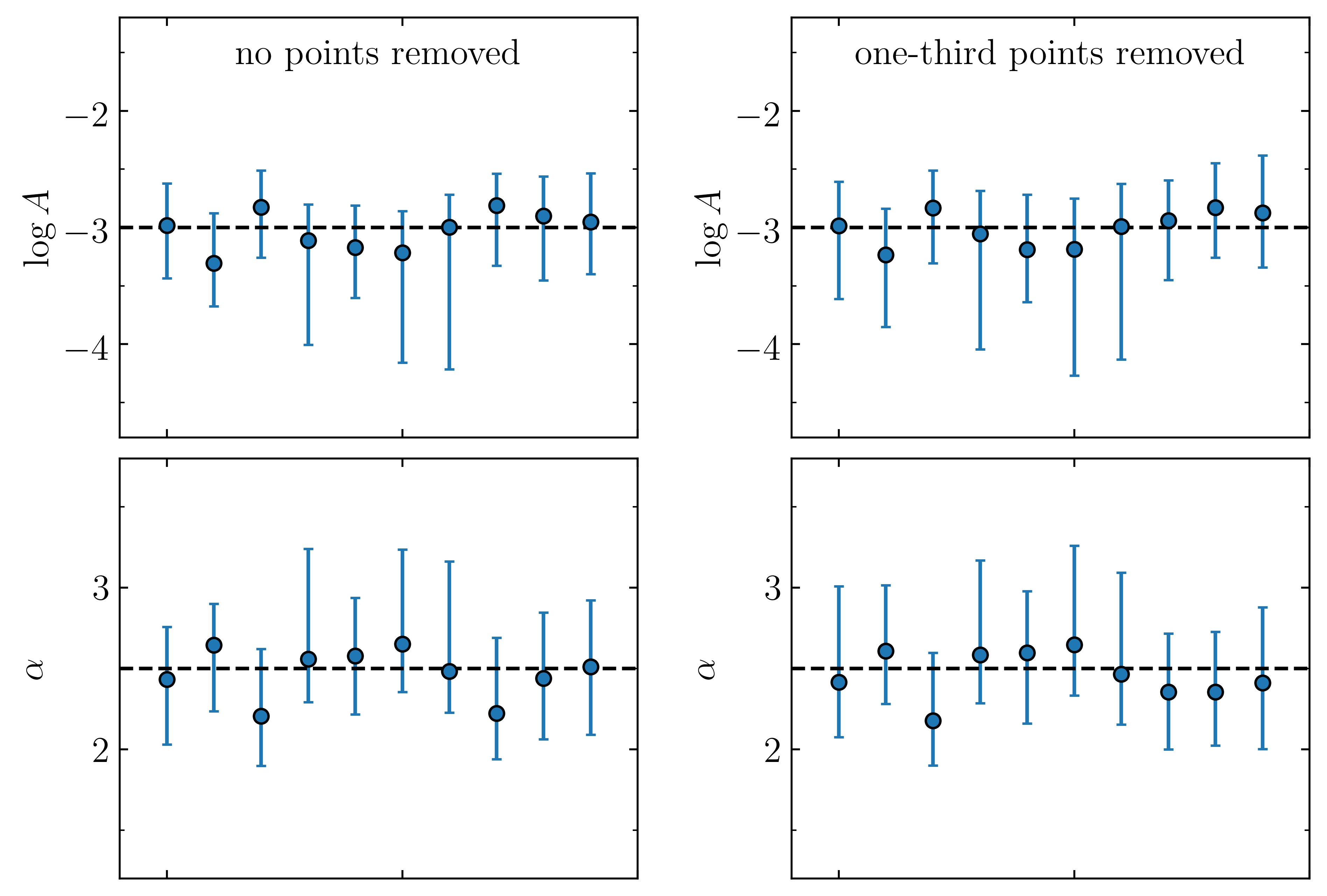}
\caption{Recovered parameters $A$ and $\alpha$ of a single power-law PSD (equation \ref{eqn_pow}) over 10 simulation tests
for cases of (left) no points removed and (right) one-third points removed.
Errorbars represents the 2$\sigma$ (95\%) errors.
Dashed lines represent the input values. The input slope $\alpha=2.5$ means that there are strong 
power leakages.}
\label{fig_sim_pow}
\end{figure}

In calculations, we first generate an evenly spaced series $\mathbi{f}$ over a frequency grid $\omega_k=k/(V W M\Delta T_{\rm sim})$ with $k=0,...,VWM/2$
using equation (\ref{eqn_freq}). With the inverse Fourier transform, we can obtain an evenly spaced time series  that has 
a time resolution of $\Delta T_{\rm sim}$ and 
a total duration of $V W M\Delta T_{\rm sim}$.
Here, $M=T/\Delta T_{\rm med}$, $T$ is the total duration of the data, and $T_{\rm med}$ is the mediate sampling interval of the data.
With these configurations, the time resolution of the generated time series is $\Delta T_{\rm sim}=\Delta T_{\rm med}/W$ 
and the total duration is $T_{\rm sim}=VT$. To take into account the effects of power leakage or aliasing, $V$ and $W$ 
should be generally large than one (\citealt{Uttley2002, Vaughan2010}). We use $V=W=2$ throughout the paper.
We finally interpolate the generated time series on the observed time points and calculate the likelihood probability in equation (\ref{eqn_post}).
Linear interpolation is sufficient provided the time resolution of the generated time series 
is smaller than the cadence of the data ($W>1$).

The prior probabilities for the parameters are assigned as follows: the priors for $\mathbi{f}$ are set to Gaussians 
according to equation (\ref{eqn_freq}); for the other parameters, if their typical value ranges are known, a uniform prior
is used; otherwise, if the parameter information is completely unknown, a logarithmic prior is used (\citealt{Sivia2006}).
For all the priors, we set a reasonably broad but still finite range to avoid the posterior impropriety (\citealt{Sivia2006}, Ch. 4).

\begin{figure*}
\centering
\includegraphics[width=0.8\textwidth, trim=0pt 10pt 0pt 10pt]{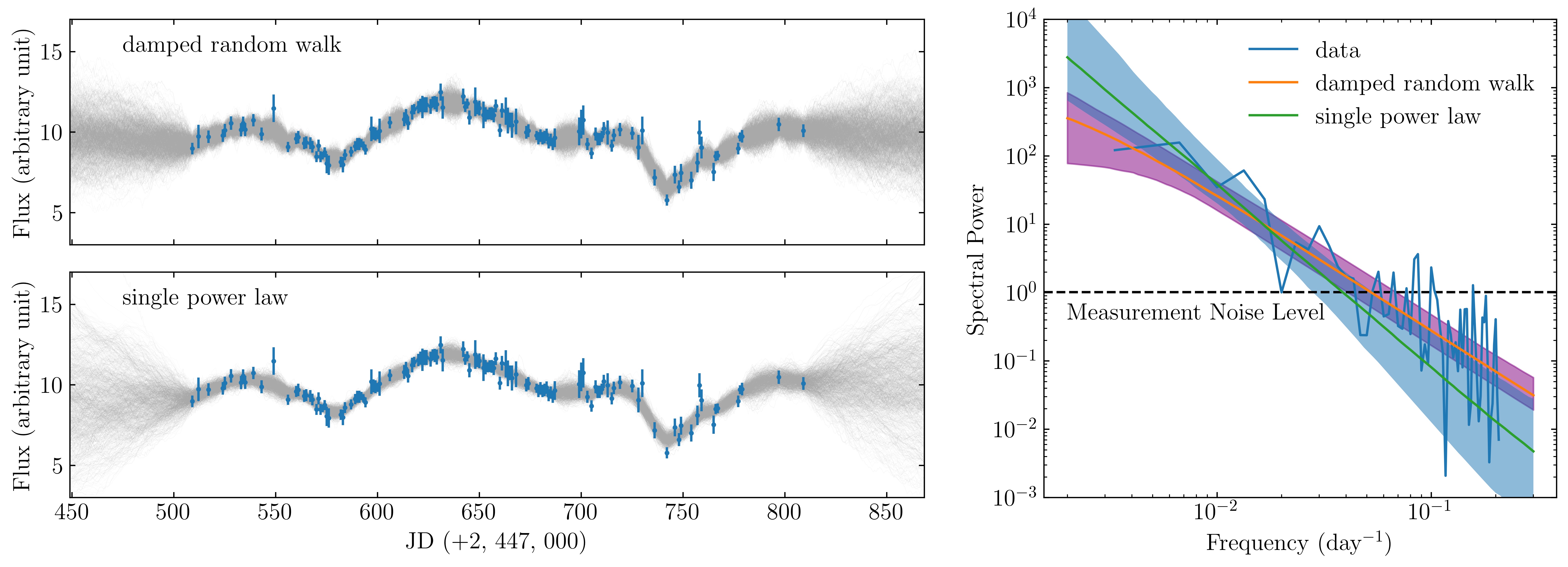}
\caption{(Left) Reconstructions for the 5100~{\AA} light curve of NGC 5548 with a damped random walk and single power-law PSD model.
(Right) The periodogram of the 5100~{\AA} light curve and the best recovered PSD for the damped random walk and single power-law models.
Shaded areas represents the 2$\sigma$ error bands (95\% confidence level).
Dashed line represents the measurement noise level. 
The periodogram is calculated by linearly interpolating the data to an even time grid.}
\label{fig_ngc5548_lc}
\end{figure*}

\subsection{Markov Chain Monte Carlo Implementation}
We use the Markov Chain Monte Carlo (MCMC) method to explore the posterior probability in equation (\ref{eqn_post})
and determine the best estimate and the associated uncertainties for the parameters.
We employ the diffusive nested sampling (DNS) algorithm proposed by \cite{Brewer2011} to construct Markov chains.
The DNS algorithm is effective at exploring multimodal distributions and strong
correlations between parameters. Another advantage of using DNS algorithm is its inherent capability of computing 
Bayesian evidence in equation (\ref{eqn_evidence}).

The fast Fourier transform algorithm (e.g., implemented by the public library \texttt{FFTW}\footnote{\url{http://fftw.org}}) 
has a time complexity of $O(N\log N)$ for a series with $N$ points. Our approach overall has the same 
time complexity of $O(N\log N)$. We implement the approach on the standardized parallel Message Passing Interface (MPI) 
to further improve the computational performance.

\subsection{Validity of the Approach}

In Fig. \ref{fig_sim}, we show one exemplary simulation test for the validity of our approach.
The input PSD is set to be a bending power law (e.g., \citealt{Uttley2002})
\begin{eqnarray}
S(\nu) = \left\{\begin{array}{cc}
          A \left(\frac{\nu}{\nu_{\rm b}}\right)^{-\alpha_{\rm hi}}+C~~~{\rm for~\nu > \nu_{\rm b}},\\
          A \left(\frac{\nu}{\nu_{\rm b}}\right)^{-\alpha_{\rm lo}}+C~~~{\rm otherwise},
         \end{array}\right.
\label{eqn_bending_pow}
\end{eqnarray}
where $A=10.0$ (arbitrary unit), $\alpha_{\rm hi}=2.5$, $\alpha_{\rm lo}=1.0$, $\nu_{\rm b}=1.5\times10^{-2}~{\rm day}^{-1}$,
and $C=0.01$ (arbitrary unit). We make the PSD break to a constant below 
$5.0\times10^{-4}{\rm~day}^{-1}$ to ensure that a physically 
meaningful PSD has to flatten for the total power to converge.
We generate a mock light curve with a cadence of 2 days and a duration of 600 days. 
The measurement noises are set to be 0.1 (arbitrary unit).
The priors for $A$, $\nu_{\rm b}$, and $C$
are set to be logarithmic and the priors for $\alpha_{\rm hi}$ and 
$\alpha_{\rm lo}$ are set to be uniform over a range $(1, 5)$ and $(-3, 4)$, respectively.
The bottom panels of Fig. \ref{fig_sim} show the posterior distributions of the PSD parameters,
which are generally consistent with the input values. We in total run 10 tests and find that all the PSD 
parameters are recovered at 2$\sigma$ level. This indicates feasibility of our approach.

The red-noise leakages are crucial for PSDs with slopes $\alpha > 2$,
which distort the slopes toward $\alpha=2$  (e.g., \citealt{Uttley2002}). Unfortunately, 
AGN PSDs typically have slopes of $\alpha=2-3$ (\citealt{Mushotzky2011, Gonzalez-Martin2012}). 
In our approach, using a much broader frequency series (i.e., $V\gtrsim 10$) can 
take account of red-noise leakages. However, this will significantly increase the computation overhead for Fourier 
transform. We instead employ the simple but effective ``end matching'' method 
to reduce leakage biases (\citealt{Fougere1985}). This method subtracts a linear trend from the time series to ensure that the first 
and end data points have equal values. As such, some fraction (but not all) of the leakages are removed.
We perform 10 simulation tests with a single power-law PSD as
\begin{eqnarray}
S(\nu) = A \nu^{-\alpha},
\label{eqn_pow}
\end{eqnarray}
where $A=1.0\times10^{-3}$ (arbitrary unit) and $\alpha=2.5$. The stochastically generated light curves 
have a cadence of 1 day and a duration of 200 days.
The measurement noises are again set to be 0.1 (arbitrary unit). The prior for $A$ is set to be logarithmic and 
for $\alpha$ is set to be uniform over a range (0, 5). The left panels of Fig. \ref{fig_sim_pow} shows 
the recovered values of $A$ and $\alpha$ with 2$\sigma$ errors for the 10 tests.
To simulate real observation data, we also randomly remove one-third points in each mock light curve and 
show the results in the right panels of Fig. \ref{fig_sim_pow}. In both cases, our approach recovers 
the PSD parameters within 2$\sigma$ uncertainties and seems fairly immune to missing data points.

\section{A Case Study: Application to NGC~5548}
We apply our approach to the 5100~{\AA} continuum light curve of NGC 5548 monitored in 1989 by
the international  AGN Watch program\footnote{\url{http://www.astronomy.ohio-state.edu/~agnwatch}} 
(\citealt{Peterson2002}). We use two PSD model to fit the data. One is a single power law with 
a form as equation (\ref{eqn_pow}); the other is a damped random walk model with a form of (e.g., \citealt{Zu2011})
\begin{eqnarray}
S(\nu) = \frac{A}{1+(\nu/\nu_{\rm d})^2},
\end{eqnarray}
where the parameters $A$ and $\nu_{\rm d}$ are related with the  typical damping time scale $\tau_{\rm d}$ and 
the long-term standard deviation $\sigma_{\rm d}$ of the random walk process through 
$\tau_{\rm d}=1/(2\pi\nu_{\rm d})$ and $\sigma_{\rm d}^2 = \pi A \nu_{\rm d}/2$. The priors for $A$ and $\nu_{\rm d}$ are  
both set to be logarithmic, and $\nu_{\rm d}$ is additionally bounded with a lower limit of $1/(2\pi T)$, where the duration of
the light curve $T=300$ days.

The best inferred values are $\log A=1.83\pm0.53$ and $\log\ (\nu_{\rm d}/{\rm day}^{-1})=-2.72\pm0.28$ for 
the damped random walk model and $\log A=-4.92\pm0.71$ and $\alpha=2.73\pm0.39$ for the single power-law model. 
The left panel of Fig.~\ref{fig_ngc5548_lc} shows the stochastic reconstructions for the light curve of NGC~5548. 
Careful inspection shows that at short time scales, the reconstructed light curves with the damped random walk model have overall
larger variations than these with the single power-law model. This is because for the single power-law model, 
low-frequency powers are dominated, which are statistically easier to produce smooth light curves. Meanwhile, 
the spectral slope $\alpha$ of the single power law is larger than $2$. In contrast, the damped random walk model
has a fixed slope of $2$ at high frequency. From the right panel of Fig. \ref{fig_ngc5548_lc}, 
we can find that the damped random walk model is inclined to match the periodogram beneath the measurement noise level, indicating
that it overfits the short-timescale variations that are probably  arisen from 
the measurement noise. Here, the measurement noise level is calculated as $S_{\rm noise} = 2\Delta T_{\rm mean} \sigma_{\rm mean}^2$, where 
$\Delta T_{\rm mean}$ and $\sigma_{\rm mean}$ are the mean sampling interval and the mean measurement error.

The calculated Bayes evidence (see equation \ref{eqn_evidence}) is $\ln P(D|S_{\rm drw})=39.31\pm0.28$ for the damped 
random walk model and  $\ln P(D|S_{\rm spl})=41.01\pm0.21$ for the single power-law model. Here the uncertainties
are estimated by running the approach 10 times and assigning the uncertainties the standard deviations.
The resulting Bayes factor is
\begin{eqnarray}
K = \frac{P(D|S_{\rm spl})}{P(D|S_{\rm drw})} =  5.47\pm1.92,
\end{eqnarray}
meaning that the damped random walk model is not preferable over the single power-law model for the data of NGC~5548
(we, however, keep in mind the possible limitations of Bayes factor; see \citealt{Gelman2004}, Ch. 6).

\begin{figure}
\centering
\includegraphics[width=0.46\textwidth, trim=0pt 20pt 0pt 0pt]{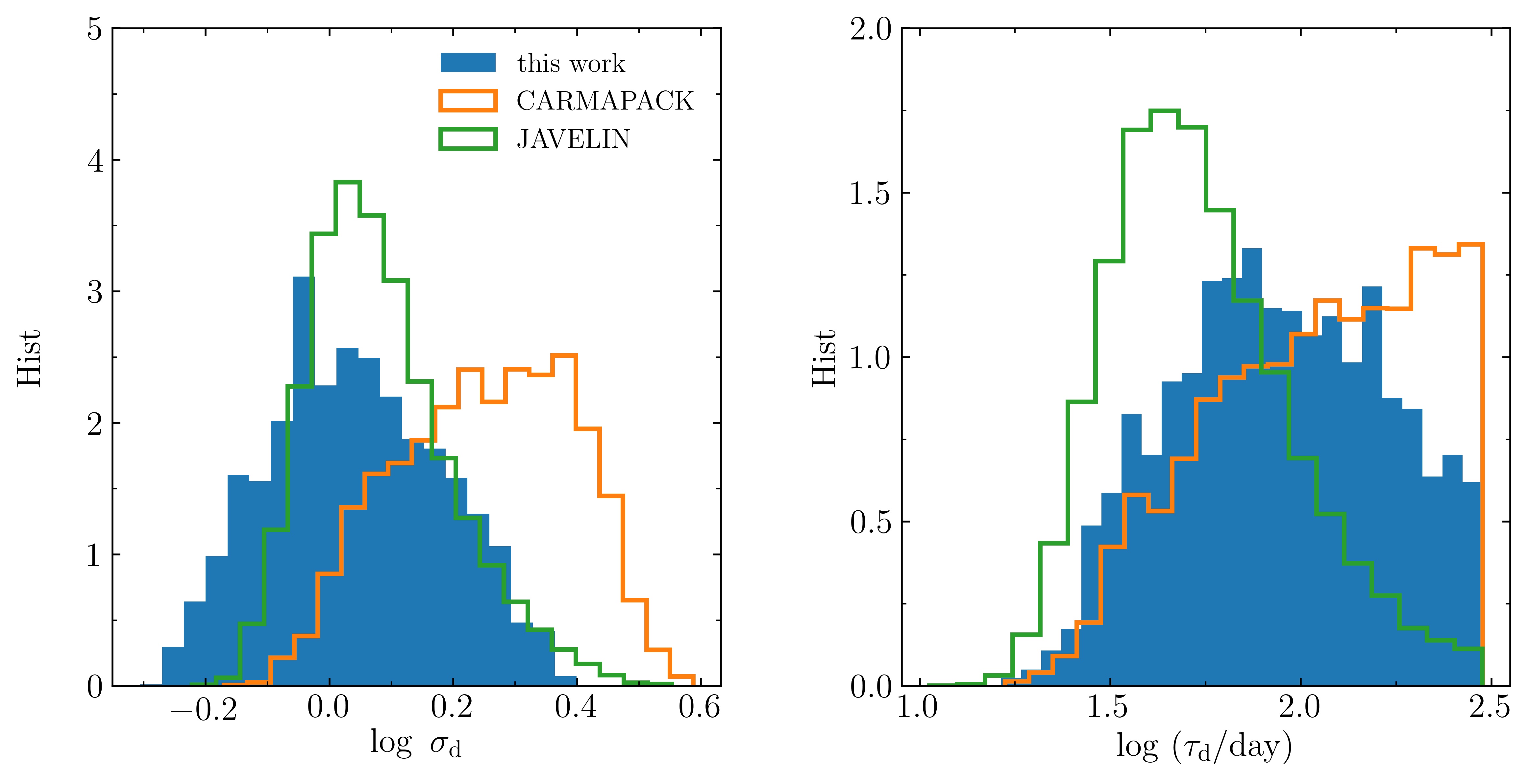}
\caption{Comparison of the posterior distributions for the parameters $\sigma_{\rm d}$ and $\tau_{\rm d}$ of 
the damped random walk model from the methods \texttt{CARMAPACK} (\citealt{Kelly2014}), \texttt{JAVELIN} (\citealt{Zu2011}), and this work
for the light curve of NGC 5548 shown in Fig.~\ref{fig_ngc5548_lc}.}
\label{fig_comp}
\end{figure}

To further test the validity of our approach, in Fig. \ref{fig_comp}, we compare the posterior distributions 
of the parameters $\sigma_{\rm d}$ and $\tau_{\rm d}$ for the damped random walk model obtained from 
\texttt{CARMAPACK}\footnote{\url{https://github.com/brandonckelly/carma_pack}.} (\citealt{Kelly2014}), 
\texttt{JAVELIN}\footnote{\url{https://bitbucket.org/nye17/javelin}.} (\citealt{Zu2011}), and our approach.
\texttt{CARMAPACK} describes light curves using a continuous-time autoregressive moving average (CARMA) process.
A CARMA process is characterized by an autoregressive order $p$ and moving average order $q$. 
The damped random walk is a particular case with $p=1$ and $q=0$.  \texttt{JAVELIN} directly employs 
the damped random walk to fit light curves in time domain. There are moderate differences in the detailed distributions 
of $\sigma_{\rm d}$ and $\tau_{\rm d}$, for example, the distributions from \texttt{CARMAPACK} tends towards 
larger values for both $\sigma_{\rm d}$ and $\tau_{\rm d}$. These differences may be ascribed to different 
analysis methods and MCMC sampling algorithms.
Nevertheless, the best estimates are consistent to within uncertainties among the three methods\footnote{However,
we note that there are possible biases on the recovery of the true values using the damped random walk model
due to the limited baseline of the data, as pointed out by \cite{Kozlowski2017b}.}.

\section{Discussion and Conclusion}
We developed a new forward approach to simultaneously measure power spectra and reconstruct 
AGN light curves. The approach models AGN variations by a series of complex Gaussian variables multiplied with 
the square root of the given PSD in frequency domain (see equation \ref{eqn_freq}) and then transforms the series back to time domain 
to fit the observed data with measurement noises. The approach is formulated in a Bayesian framework
with the capability of comparing different PSD models through Bayesian model selection.
The parameters for the PSD models and light curves and their associated uncertainties are determined by the MCMC technique 
with an advanced nested sampling algorithm, which is apt to calculate the Bayes evidence and significantly facilitates 
the selection of PSD models. Application of our approach to the 5100~{\AA} light curve of NGC~5548
shows that the widely used damped random walk model is not preferable over the single power-law model.

The caveats of our new approach include: First, the time consumption of implementing the Fourier transform is relatively high. 
Although the well-developed fast Fourier transform algorithm improves the computation speed with a complexity of $O(N\log N)$, 
MCMC sampling is still computationally expensive. This can be overcome to some extent by using parallel computation.
Second, our approach is not optimal for light curves with fine sampling separated by large gaps, as this requires 
very dense and large series to describe the whole light curves. A possible way around this issue 
is splitting the light curves into segments and use separated frequency series to model them (but performing MCMC 
inference as a whole). Third, there is evidence that variations from accreting black holes (AGNs and X-ray 
binaries) commonly show the linear rms-flux relation (e.g., \citealt{Uttley2005} and references therein), which implies 
that the underlying variability processes are multiplicative. \cite{Uttley2005} demonstrated that a simple 
exponential transform of a linear time series can fully explain the rms-flux relation. It is easy to 
include such a transform in our approach. However, this will also modify the output PSD shape relative to the input one,
therefore, we do not include this exponential transform in the present paper.

We have developed software to implement our approach. The software is written in C language with the standardized
MPI so that it is portable to a wide range of computers/supercomputer clusters. The software is available at
\url{https://github.com/LiyrAstroph/RECON}.

\section*{Acknowledgments}
We thank the referee for useful suggestions that improved the manuscript.
This research is supported in part by the National Key R\&D Program of China (2016YFA0400700), 
by the CAS Key Research Program (KJZDEW-M06), and 
by grant No. 1113006, 11690024, and U1431228 from the National Natural Science Foundation of China (NSFC).
Y.R.L. acknowledges financial support from the NSFC grant No. 11573026 
and from the CAS Strategic Priority Research Program grant No. XDB23000000.

\bsp	
\label{lastpage}

\begin{thebibliography}{99}
\bibitem[\protect\citeauthoryear{Brewer et al.}{2011}]{Brewer2011}
  Brewer B.~J., P\'{a}atay L. B, \& Cs\'{a}nyi G.\ 2011, Stat. Comput., 21, 649
\bibitem[\protect\citeauthoryear{Done et al.}{1992}]{Done1992} 
  Done C., Madejski G.~M., Mushotzky R.~F., Turner T.~J., Koyama K., Kunieda H., 1992, ApJ, 400, 138 
\bibitem[\protect\citeauthoryear{Drake et al.}{2009}]{Drak2009} 
  Drake A.~J., et al., 2009, ApJ, 696, 870 
\bibitem[Fougere(1985)]{Fougere1985} 
  Fougere, P.~F.\ 1985, \jgr, 90, 4355 
\bibitem[\protect\citeauthoryear{Gelman et al.}{2004}]{Gelman2004}
  Gelman A., Carlin J. B., Stern H. S., \& Rubin, D. B., 2004, Bayesian Data Analysis, 2nd edn, Chapman \& Hall, London
\bibitem[\protect\citeauthoryear{Gonz{\'a}lez-Mart{\'{\i}}n \& Vaughan}{2012}]{Gonzalez-Martin2012} 
  Gonz{\'a}lez-Mart{\'{\i}}n O., Vaughan S., 2012, A\&A, 544, A80 
\bibitem[\protect\citeauthoryear{Ivezic et al.}{2008}]{Ivezic2008} 
  Ivezic Z., et al., 2008, arXiv, arXiv:0805.2366 
\bibitem[\protect\citeauthoryear{Kelly, Bechtold, \& Siemiginowska}{2009}]{Kelly2009} 
  Kelly B.~C., Bechtold J., Siemiginowska A., 2009, ApJ, 698, 895-910 
\bibitem[Kelly et al.(2014)]{Kelly2014} 
  Kelly, B.~C., Becker, A.~C., Sobolewska, M., Siemiginowska, A., \& Uttley, P.\ 2014, \apj, 788, 33 
\bibitem[\protect\citeauthoryear{Kochanek et al.}{2017}]{Kochanek2017} 
  Kochanek C.~S., et al., 2017, PASP, 129, 104502 
\bibitem[\protect\citeauthoryear{Koz{\l}owski}{2017a}]{Kozlowski2017a} 
  Koz{\l}owski S., 2017a, ApJ, 835, 250 
\bibitem[\protect\citeauthoryear{Koz{\l}owski}{2017b}]{Kozlowski2017b} 
  Koz{\l}owski S., 2017b, A\&A, 597, A128 
\bibitem[\protect\citeauthoryear{Li et al.}{2014}]{Li2014} 
  Li Y.-R., Wang J.-M., Hu C., Du P., Bai J.-M., 2014, ApJ, 786, L6 
\bibitem[Li et al.(2013)]{Li2013} 
  Li Y.-R., Wang J.-M., Ho L.~C., Du P., \& Bai J.-M., 2013, \apj, 779, 110
\bibitem[\protect\citeauthoryear{Mushotzky et al.}{2011}]{Mushotzky2011} 
  Mushotzky R.~F., Edelson R., Baumgartner W., Gandhi P., 2011, ApJ, 743, L12 
\bibitem[\protect\citeauthoryear{Pancoast, Brewer, \& Treu}{2014}]{Pancoast2014} 
  Pancoast A., Brewer B.~J., Treu T., 2014, MNRAS, 445, 3055 
\bibitem[\protect\citeauthoryear{Peterson et al.}{2002}]{Peterson2002} 
  Peterson B.~M., et al., 2002, ApJ, 581, 197 
\bibitem[Priestley (1981)]{Priestley1981}
  Priestley M. B., 1981, Spectral Analysis and Time Series. Academic Press, London, p. 389
\bibitem[\protect\citeauthoryear{Shakura \& Sunyaev}{1973}]{Shakura1973} 
  Shakura N.~I., Sunyaev R.~A., 1973, A\&A, 24, 337 
\bibitem[\protect\citeauthoryear{Sivia \& Skilling}{2006}]{Sivia2006}
  Sivia D. S., Skilling J., 2006, Data Analysis: A Bayesian Tutorial, 2nd edn. Oxford Univ. Press, Oxford
\bibitem[\protect\citeauthoryear{Timmer \& Koenig}{1995}]{Timmer1995} 
  Timmer J., Koenig M., 1995, \aap, 300, 707 
\bibitem[\protect\citeauthoryear{Ulrich, Maraschi, \& Urry}{1997}]{Ulrich1997} 
  Ulrich M.-H., Maraschi L., Urry C.~M., 1997, ARA\&A, 35, 445 
\bibitem[\protect\citeauthoryear{Uttley, McHardy, \& Papadakis}{2002}]{Uttley2002} 
  Uttley P., McHardy I.~M., Papadakis I.~E., 2002, MNRAS, 332, 231 
\bibitem[\protect\citeauthoryear{Uttley, McHardy, \& Vaughan}{2005}]{Uttley2005} 
  Uttley P., McHardy I.~M., Vaughan S., 2005, MNRAS, 359, 345 
\bibitem[\protect\citeauthoryear{Vaughan}{2010}]{Vaughan2010} 
  Vaughan S., 2010, MNRAS, 402, 307 
\bibitem[\protect\citeauthoryear{Zhu \& Xue}{2016}]{Zhu2016} 
  Zhu S.~F., Xue Y.~Q., 2016, ApJ, 825, 56 
\bibitem[\protect\citeauthoryear{Zu, Kochanek, \& Peterson}{2011}]{Zu2011} 
  Zu Y., Kochanek C.~S., Peterson B.~M., 2011, ApJ, 735, 80 
\end{thebibliography}
\end{document}